\documentclass[aps,preprint,preprintnumbers,amsmath,amssymb]{revtex4}
\usepackage{amsmath,mathrsfs,amsbsy,color,graphicx,bm,amsthm,amsfonts}
\usepackage{units}
\usepackage{bbm}
\usepackage{times}
\usepackage{dcolumn}
\usepackage{mathrsfs}% Align table columns on decimal point
\usepackage{amsmath,amssymb,epsfig}
\usepackage{hyperref}
\begin{document}

\title{Estimating the Unruh effect via  entangled many-body probes}

\author{Jieci Wang, Li Zhang,  Songbai Chen\footnote{csb3752@hunnu.edu.cn}, and Jiliang Jing\footnote{jljing@hunnu.edu.cn}}
\affiliation{Department of Physics, and Key Laboratory of Low
Dimensional \\Quantum Structures and Quantum
Control of Ministry of Education,\\
 Hunan Normal University, Changsha, Hunan 410081, China}.

\begin{abstract}
 We study the estimation of parameters in a quantum metrology scheme based on entangled many-body Unruh-DeWitt  detectors.  It is found that the precision for the estimation of Unruh effect can be enhanced via initial state preparations and  parameter selections.   It is shown that  the precision in the estimation of the Unruh temperature in terms of a many-body-probe metrology is always better than the precision
in two probe strategies. The proper acceleration for Bob's detector and the interaction between the accelerated detector and the external field have significant influences on the precision for the Unruh effect's estimation. In addition, the probe state prepared with  more excited atoms in the initial state is found to perform better than less  excited  initial states. However, different from the estimation of the Unruh temperature,  the estimation of the effective coupling parameter for the accelerated detector requires  more total atoms but  less excited atoms in the estimations.

 \vspace*{0.2cm}
\end{abstract}

\maketitle

\section{introduction}

  As predicted by quantum field theory in curved spacetime, quantum fluctuation will produce  a local change in
energy of an Unruh-DeWitt detector \cite{UW84}. It was found that  the detectors will be
thermalized at a temperature defined by some characteristic inverse length scale of the spacetime or motion through
them, e.g. acceleration \cite{Matsas01, Unruh,Unruhr}, surface gravity \cite{Hawking74}, or Hubble constant \cite{Hawking77}.  The feature that all three of these effects have in common is the existence of
an event horizon. The  detector will observe a radiation  originating from quantum fluctuations near the horizon.  The Unruh effect  \cite{Matsas01, Unruh,Unruhr} predicts the thermality of  a uniformly accelerated detector in the Minkowski vacuum. This thermality can be demonstrated by tracing the field modes beyond the Rindler event horizon  therefore manifests itself as  a  decoherence-like effect. The Unruh-Hawking effect also makes a deep connection between  black hole thermodynamics and  quantum physics, such as the understanding of entanglement entropy and quantum nonlocality in curved spacetime.

Despite its crucial role in  physics,  the experimental detection  of the Unruh effect is  an open program on date. The main technical obstacle is that the Unruh temperature is smaller than 1 Kelvin even for  accelerations up to $10^{21} m/s^2$  \cite{unruhreview}.  This means the detectable  Unruh temperature  lies far below the observable threshold with the experimentally achievable acceleration.  Since experimental detection of the Unruh radiation is too difficult,  people turn sight to  the  easier but still conceptually rich studies on the simulation \cite{natureobhawking} and estimation of this effect.
We know that  quantum metrology  aims to improve the precision in estimating parameters  via quantum  strategies \cite{Giovannetti2011}. The estimation is based on measurements made on a probe system that undergoes an
evolution depending on the estimated parameters. Recently,  quantum metrology  has been applied to enhance the  detection of gravitational wave beyond the standard quantum limit \cite{Abbott}, the exploration of the Earth's Schwarzschild parameters \cite{DEB,wangmetro,earthmetro},  and the estimation of cosmological  parameters in expanding universe \cite{wangmetro2,xiaobao}.  In particular, researchers found that quantum strategies can be employed to enhance the estimation of the Unruh-Hawking effect, both for accelerated free modes \cite{aspachs,HoslerKok2013,RQI6,wangmetro3,unruhEPJC}, local modes in moving cavities \cite{RQM,RQM2}, and accelerated  detectors \cite{wangmetro1,unruhJHEP2,unruhJHEPun1}. However, it is worthy of note that all probe states in the above-mentioned quantum enhanced estimation tasks for the Unruh-Hawking effect are prepared in bipartite entangled states.

 In this paper, we employ a multipartite entangled probe strategy to estimate the Unruh temperature and related parameters. The probes for the quantum metrology are prepared by  $n$ entangled Unruh-DeWitt detectors. Each detector is modeled by a two-level atom which interacts only with the neighbor field in the Minkowski vacuum \cite{UW84}. We assume that the second atom of multiparty system, carried by Bob, moves with constant acceleration and interacts with a massless scalar field  while other  atoms  keep stationary. To analyze the maximum achievable precision in the estimation of the parameters, we calculate  quantum Fisher information  with respect to them.  It is worth mentioning  that, like the quantum Fisher information, the  Wigner-Yanase-Dyson skew information \cite{skewinfor} is also a variant of the Fisher information in the quantum regime. The latter is related to the quantum Hellinger distance \cite {Luo04} and has been shown to satisfy some nice properties relevant to quantum coherence \cite{Yadin16,unruhEPJC}.  Recently, the authors of  \cite{unruhEPJC}  developed a Bloch vector representation of the Unruh channel  and made a comparative study on the quantum Fisher and Skew information for free Dirac field modes. In this paper we only study the quantum Fisher information because the quantum Fisher and Skew information are in fact two different aspects of the classical Fisher information in the quantum regime \cite{Petz}.

 The outline of the paper is organized as follows.
  In Section.~\ref{sec:Evomulti}, the evolution of the multiparty system is presented where one of the detectors  travels with uniform acceleration. In Sec.~\ref{sec:Fisher}, we start with introducing some key concepts for quantum metrology  especially the quantum Fisher information. Then we analyze the  quantum Fisher information for estimating  Unruh temperature $T$ and the effective coupling parameter $\nu$.
  The Sec.~\ref{sec:conclusion} is devoted to a brief summary.

\section{The evolution of  multiparty quantum system with an accelerated atom}
\label{sec:Evomulti}
In this section, we briefly introduce the dynamics of the  multiparty entangled Unruh-DeWitt detectors. Assuming that the probe system consists of $n$ atoms,  the  initial state is prepared in a symmetric $Z$-type multipartite state \cite{Ben-Av,Lizhang}
\begin{eqnarray}\label{Multiple1}
 |\psi_{t_{0}}\rangle
 & = & \sqrt{\frac{(n-k)!k!}{n!}}(|111...000\rangle\nonumber\\
 & + & |11...01...00\rangle+...+|000...111\rangle),
\end{eqnarray}
where $k$ atoms own the excited energy eigenstate $|1\rangle$, while the rest $n-k$ atoms lie in the ground state $|0\rangle$. If $k=1$, it degenerates into a symmetric $W$-type state.

The total Hamiltonian of the entire probe system is
\begin{equation}\label{Multipleh}
  H_{n\phi}=\sum_{i=1}^{n}H_{i}+H_{KG}+H_{int}^{B\phi},
\end{equation}
where $H_{KG}$ stands for the Hamiltonian of the massless scalar field satisfying the K-G equation $\Box\phi=0$. In Eq. (\ref{Multipleh}),
$H_{i}=\Omega D_{i}^{\dagger} D_{i},i=1,2...n$ are the Hamiltonian of each atom, where $D_{i}$ and $D_{i}^{\dagger}$ represent the creation and annihilation operators of the $i$th atom, respectively. We assume that the second atom of the multiparty system, carried by the observer Bob, is uniformly accelerated for a duration time $\Delta$, while other atoms  keep static and have no interaction with the scalar field. The world line of  Bob's detector is given by
\begin{equation}
 t(\tau)=a^{-1}\sinh a \tau, x(\tau)=a^{-1}\cosh a \tau, y(\tau)=z(\tau)=0,
\end{equation}
where  $a$ is the proper acceleration of Bob.
The interaction Hamiltonian $H_{int}^{B\phi}$ between Bob's detector and the field is \cite{Landulfo,Landulfo1,wang3}
\begin{equation}\label{hamitonian}
H^{B\phi}_{int}(t)=
\epsilon(t) \int_{\Sigma_t} d^3 {\bf x} \sqrt{-g} \phi(x) [\chi({\bf x})D_{2} +
                           \overline{\chi}({\bf x})D_{2}^{\dagger}],
\end{equation}
where $ \phi(x) $ is the scalar field operator, $\epsilon(t)$ is the coupling constant, $g_{ab}$ is the Minkowski metric and $g\equiv {\rm det} (g_{ab})$. In Eq. (\ref{hamitonian}),  $ \Sigma_t $ represents the integration takes place over the global spacelike Cauchy surface. The function $\chi(\mathbf{x})$ vanishes outside a small volume
around the detector, which describes that the detector only interacts with the neighbor field.

For convenience, introducing a compact support complex function $ f(x)=\epsilon(t)e^{-i\Omega t}\chi({\bf x})$, we have $\phi(x)f\equiv Rf-Af$, where $A$ and $R$ are the advanced and retarded Green functions. Then one obtains \cite{Landulfo,Landulfo1,wang3}
\begin{equation}\label{phif}
\phi(f)\equiv \int d^4 x \sqrt{-g}\phi(x)f=i[a_{RI}(\Gamma_{-}^{*})-a_{RI}^{\dagger}(\Gamma_{+})],
\end{equation}
where  $\Gamma_{-}$ and $\Gamma_{+}$  represent the negative and positive frequency parts of $\phi(f)$ respectively, and  $a^{\dagger}_ {RI}$ and $a_{RI}$ are Rindler creation and annihilation operators in  region $I$ of Rindler spacetime. Since $\epsilon(t)$ is a roughly constant for $\Delta\gg\Omega^{-1}$, the test function $f$ approximately owns the positive-frequency part, which means $\Gamma_{-}\approx0$. And if we define $\lambda\equiv -\Gamma_{+}$, Eq. (\ref{phif}) is found to be $\phi(f) \approx i a^{\dagger}(\lambda)$.

The whole initial state of $n$-party systems and the massless scalar field is $|\Psi_{t_{0}}^{n\phi}\rangle=|\psi_{t_{0}}\rangle\otimes|0\rangle_{M}$, where $|0\rangle_{M}$ stands for Minkowski vacuum.
\iffalse Bob's  creation and annihilation operators have been donated by $D_{2}^{\dagger}$ and $D_{2}$, respectively.\fi
Here we only consider the first order under the weak-coupling limit. Using Eq. (\ref{hamitonian}) and Eq. (\ref{phif}), we can calculate the final state of the probe state  at time $t>t_{0}+\Delta$, which is
\begin{eqnarray}\label{evolvet}
|\Psi_{t}^{n\phi}\rangle
& = & T\exp[-i\int d^4 x\sqrt{-g}\phi(x)(fD_{2}+\overline{f}D_{2}^{\dagger})]|\Psi_{t_{0}}^{n\phi}\rangle \nonumber\\
& \approx & 1-i\int d^4 x\sqrt{-g}\phi(x)[fD_{2}+\overline{f}D_{2}^{\dagger}]|\Psi_{t_{0}}^{n\phi}\rangle \nonumber\\
&=&(1+a_{RI}^{\dagger}(\lambda)D_{2}-a_{RI}(\overline{\lambda})D_{2}^{\dagger})|\Psi_{t_{0}}^{n\phi}\rangle,
\end{eqnarray}
in the interaction picture, where $T$ is the time-order operator.

As discussed in \cite{Landulfo,Landulfo1, wang3,wangmetro1}, the Bogliubov transformations between the  operators of Minkowski modes and Rindler modes are
\begin{eqnarray}
a_{R I}(\overline{\lambda})&=&
\frac{a_M(\overline{F_{1 \Omega}})+
e^{-\pi \Omega/a} a_M ^{\dagger} (F_{2 \Omega})}{(1- e^{-2\pi\Omega/a})^{{1}/{2}}},
\label{aniq} \\
a^{\dagger}_{R I}(\lambda)&=&
\frac{a^{\dagger}_M (F_{1 \Omega}) +
e^{-\pi \Omega/a}a_M(\overline{F_{2 \Omega}})}{(1- e^{-2\pi\Omega/a})^{{1}/{2}}}
\label{cria},
\end{eqnarray}
where
$F_{1 \Omega}=
\frac{\lambda+ e^{-\pi\Omega/a}\lambda\circ w}{(1- e^{-2\pi\Omega/a})^{{1}/{2}}}$, and
$F_{2 \Omega}=
\frac{\overline{\lambda\circ w}+ e^{-\pi\Omega/a}\overline{\lambda}}{(1- e^{-2\pi\Omega/a})^{{1}/{2}}}$.
In $F_{1 \Omega}$ and $F_{2 \Omega}$, $w(t, x)=(-t, -x)$
denotes the wedge reflection isometry, which makes a reflection from $\lambda$ in  Rindler region $I$  to $\lambda\circ w$ in  Rindler region $II$.

By using the Bogliubov transformations given in Eqs. (\ref{aniq}-\ref{cria}),  Eq. (\ref{evolvet})  can be rewritten as
\begin{eqnarray}
|\Psi_{t}^{n\phi}\rangle
& = & |\Psi_{t_{0}}^{n\phi}\rangle +\frac{1}{(1-q)^{1/2}}\sqrt{\frac{(n-k)!k!}{n!}}
 ( |\Psi_{1}^{n}\rangle\nonumber\\
 &\otimes& q^{1/2}|1_{F_{2\Omega}}\rangle
 +  |\Psi_{2}^{n}\rangle\otimes|1_{F_{1\Omega}}\rangle\rangle),
\label{Wholefistate}
\end{eqnarray}
where the parameterized acceleration $q \equiv e^{-2\pi\Omega / a}$ has been introduced.
\iffalse where we have introduced  the parameterized acceleration $q \equiv e^{-2\pi\Omega / a}$.\fi In Eq. (\ref{Wholefistate}),
\begin{equation}
|\Psi_{1}^{n}\rangle=(\underbrace{|11...00\rangle+...+|01...11\rangle}_{C_{n-1}^{k}\;\;terms}),
\end{equation}
where Bob's atom  and $\frac{(n-1)!}{(n-k-1)!k!}$ atoms of  the rest atoms are sure to lie in excited energy eigenstates $|1\rangle$.
\begin{equation}
|\Psi_{2}^{n}\rangle=(\underbrace{|10...00\rangle+...+|00...11\rangle}_{C_{n-1}^{k-1}\;\;terms}).
\end{equation}
This means Bob's atom  is certain to be in ground state $|0\rangle$
and $\frac{(n-1)!}{(n-k)!(k-1)!}$ atoms in  the rest atoms are sure to be involved in excited energy eigenstates.

Since we only concern about the probe state after the acceleration of Bob, the degrees of freedom of the external scalar field should be  traced out.
By doing this we obtain the final density matrix of the many-body probe system
\begin{eqnarray}\label{Multiple2}
  \rho_{t}^{n}
  &=& |C|^{-2}\{|\psi_{t_{0}}\rangle\langle\psi_{t_{0}}|+\frac{n!}{(n-k)!k!}\frac{\nu^{2}}{1-q}\nonumber\\
   &\times&(q|\Psi_{1}^{n}\rangle\langle\Psi_{1}^{n}| +|\Psi_{2}^{n}\rangle\langle\Psi_{2}^{n}|)\},
\end{eqnarray}
with the energy gap $\Omega$  and the coupling constant $\epsilon$, where
$\nu^{2}\equiv\frac{\epsilon^{2}\Omega\Delta}{2\pi}%
e^{-\Omega^{2}\kappa^{2}}$ is  the effective coupling and $C=(1+\frac{q\nu^{2}(n-k)+\nu^{2}k}{(1-q)n})^{1/2}$ normalizes the final state $\rho_{t}^{n} $.

\section{Quantum metrology and quantum Fisher information for the Unruh effect}
\label{sec:Fisher}
As an important quantity in the geometry of Hilbert spaces, the quantum Fisher information has a significant impact on quantum metrology, which evaluates the state sensitivity with the perturbation of the parameter  \cite{Giovannetti2011}.
For a statistic nature, the maximum achievable precision of quantum metrology is determined
 the quantum  Cram\'{e}r-Rao bound \cite{Cramer:Methods1946}, which demands a fundamental lower bound for the covariance matrix of the estimation \cite{Braunstein1994,Braunstein1996}
\begin{eqnarray}
Var(\epsilon)\geq \frac{1}{N\mathcal{F}_{\xi}(\epsilon)},
\end{eqnarray}
 where $N$ is  the number of measurements and $\mathcal{F}_{\xi}(\epsilon)$ is the Fisher information  \cite{Braunstein1994,Braunstein1996}
 \begin{eqnarray}
\mathcal{F}_{\xi}(\epsilon)=\int \mathcal{L}_\epsilon(\frac{ \partial }{\partial\theta}\ln \ln L_\epsilon)dx.
\end{eqnarray}
Here the symmetric logarithmic derivative (SLD) Hermitian operator $\mathcal{L}_\epsilon$ is defined as $
  \partial_{\epsilon}\,\rho_{\epsilon}=\frac{1}{2}\left\{\rho_\epsilon,\mathcal{L}_\epsilon\right\}$,
where $\partial_{\epsilon}\equiv\frac{\partial}{\partial\epsilon}$
and $\left\{\,\cdot\,,\,\cdot\,\right\}$
denotes the anticommutator.
For any given POVM $\{\Pi_\xi\}$,  Fisher information establish the
bound on precision. To obtain the ultimate bounds on
precision, one should  maximize the  Fisher information over all
the possible  measurements.
Then we have  \cite{Braunstein1994, Braunstein1996}
\begin{align}
\mathcal{F}_{\xi}(\epsilon)  & \leq
\Sigma_\xi \left|\frac{Tr\left[\rho_{\epsilon} \Pi_\xi
L_T\right]}{\sqrt{Tr[\rho_{\epsilon} \Pi_\xi]}}\right|^2\nonumber
\\
& \leq \ \Sigma_\xi Tr\left[\Pi_\xi \mathcal{L}_\epsilon \rho_{\epsilon} \mathcal{L}_\epsilon\right]\nonumber
\\
& =\mathcal{F}_{Q}(\epsilon),
\end{align}
where $\mathcal{F}_{Q}(\epsilon)$ is the  quantum Fisher information \cite{Braunstein1994, Braunstein1996}
\begin{align}
\mathcal{F}_{Q}(\epsilon)
= Tr[\partial_\epsilon \rho_{\epsilon} \mathcal{L}_\epsilon]= Tr[\rho_{\epsilon} \mathcal{L}_\epsilon^2].
\label{eq:QuantumFisher}
\end{align}
Thus,  optimizing over all the possible measurements leads to  an lower quantum Cram\'{e}r-Rao bound~\cite{Braunstein1994}, i.e.,
\begin{equation}\label{Cramer-Rao}
 Var(T)\geq\frac{1}{n\mathcal{F}_{\xi}(\epsilon)}\geq \frac{1}{n\mathcal{F}_{Q}(\epsilon)}.
\end{equation}
By diagonalizing the density matrix as  $\rho_{\epsilon}=\sum_{i=1}^N \lambda_i|\psi_i\rangle\langle\psi_i|$,
the SLD operator  $\mathcal{L}_\epsilon$ owns the form
\begin{equation}
 \mathcal{L}_\epsilon=2\sum_{m,n}^N\frac{\langle\psi_k|
\partial_{\epsilon}\rho_{\epsilon}|\psi_k\rangle}{\lambda_m+\lambda_n}|\psi_n\rangle\langle\psi_m|,
\end{equation}
and the quantum Fisher information can be obtained by \cite{Zhong}
\begin{equation}\label{mixed}
\mathcal{F}_{Q}(\epsilon)=2\sum_{m,n}^N\frac{|\langle\psi_m|
\partial_{\epsilon}\rho_{\epsilon}|\psi_n\rangle|^2}{\lambda_m+\lambda_n},
\end{equation}
where the eigenvalues $\lambda_i\geq0$ and $\sum_i^N\lambda_i=1$.

Then we   calculate the  quantum Fisher information of Unruh temperature $T$  for the final  probe state  to find the optimal choice to estimate the temperature. Obviously, the final density matrix Eq. (\ref{Multiple2}) of the probe state is not full matrix because its rows with the basis $|000...000\rangle$ or
$|111...111\rangle$ are zero.
 According to Eqs.  (\ref{Multiple2}) and  (\ref{mixed}), we can calculate the nonzero eigenvalues
 \begin{eqnarray*}
   \Gamma_{1} &=& \frac{(1-q)n}{(1-q)n+qv^2(n-k)+v^2k},\\
  \Gamma_{2} &=&  \frac{qv^2(n-k)}{(1-q)n+qv^2(n-k)+v^2k}, \\
   \Gamma_{3} &=& \frac{v^2k}{(1-q)n+qv^2(n-k)+v^2k}
 \end{eqnarray*}
as well as the  quantum Fisher information of Unruh temperature $T$  for the final  probe state. The  quantum Fisher information  is found to be
\begin{equation}
  \mathcal{F}_{Q}(T)=\frac{\Omega^{2}qv^{2}[n^{2}-(1-q)k^{2}v^{2}+(1-q)kn(v^{2}-1)]}{(1-q)T^{4}[(1-q)kv^{2}+n(1-q+qv^{2})]^{2}}.\label{quantum Fisher informationT}
\end{equation}

\begin{figure}
\begin{center}
\includegraphics[height=0.5\textheight]{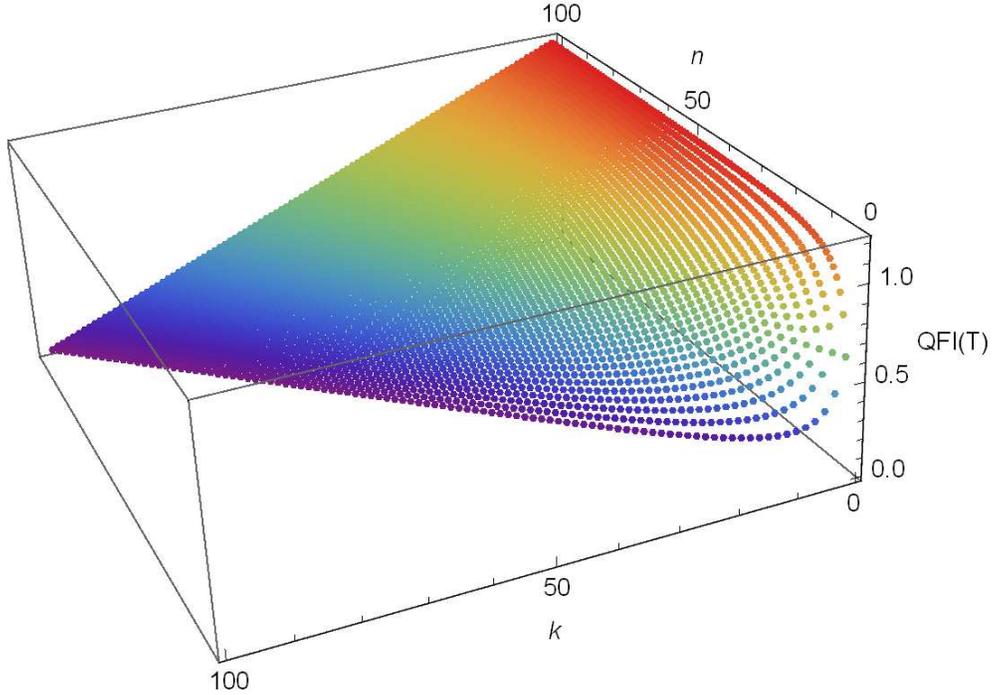}
\caption{The quantum Fisher information for estimating Unruh temperature $T$ as functions of the total atoms $n$ and the excited atoms $k$ in the initial probe state Eq. (1). The energy gap $\Omega$, the acceleration $a$ and the  effective coupling parameter $\nu$ are fixed as $\Omega=0.2$, $a=0.3$ and $\nu=0.1$, respectively. }
\label{QFITnk}
\end{center}
\end{figure}

In Fig. \ref{QFITnk}, we plot the quantum Fisher information for estimating the Unruh temperature $T$ as a function of the total atoms $n$ and the excited atoms $k$ in the initial probe state. \iffalse the energy gap $\Omega$ should be small enough  to make the atom easier to be excited and de-excited with measurements more than once. \fi  In this model, the value of effective coupling parameter $\nu$ should be  small enough  to keep the perturbative approach valid for large times. It is shown that the amount of  the total atoms $n$ and the excited atoms $k$  have significant influences on the value of quantum Fisher information.
We find that the probe system owning the more total atoms would gain the higher quantum Fisher information. This means that the precision in the estimation of the Unruh temperature in terms of a many-body probe state is always better than precision
in a bipartite probe system.  That is to say, compared with previous bipartite metrology proposals, the  multiparty-entangled-probe  proposal  for the quantum metrology of the Unruh effect  is more workable and reliable.
 However, it is shown that with same atoms in the initial probe state, the less the excited atoms, the larger the quantum Fisher information. This means the number of excited atoms is disadvantage for the estimation of Unruh temperature. Therefore, the initial probe state prepared in $W$-type  state  always performs  better than the $Z$-type probe state  (if $k\neq1$) for the same size initial state in the estimation of  temperature $T$.

\begin{figure}
\begin{center}
\includegraphics[height=0.5\textheight]{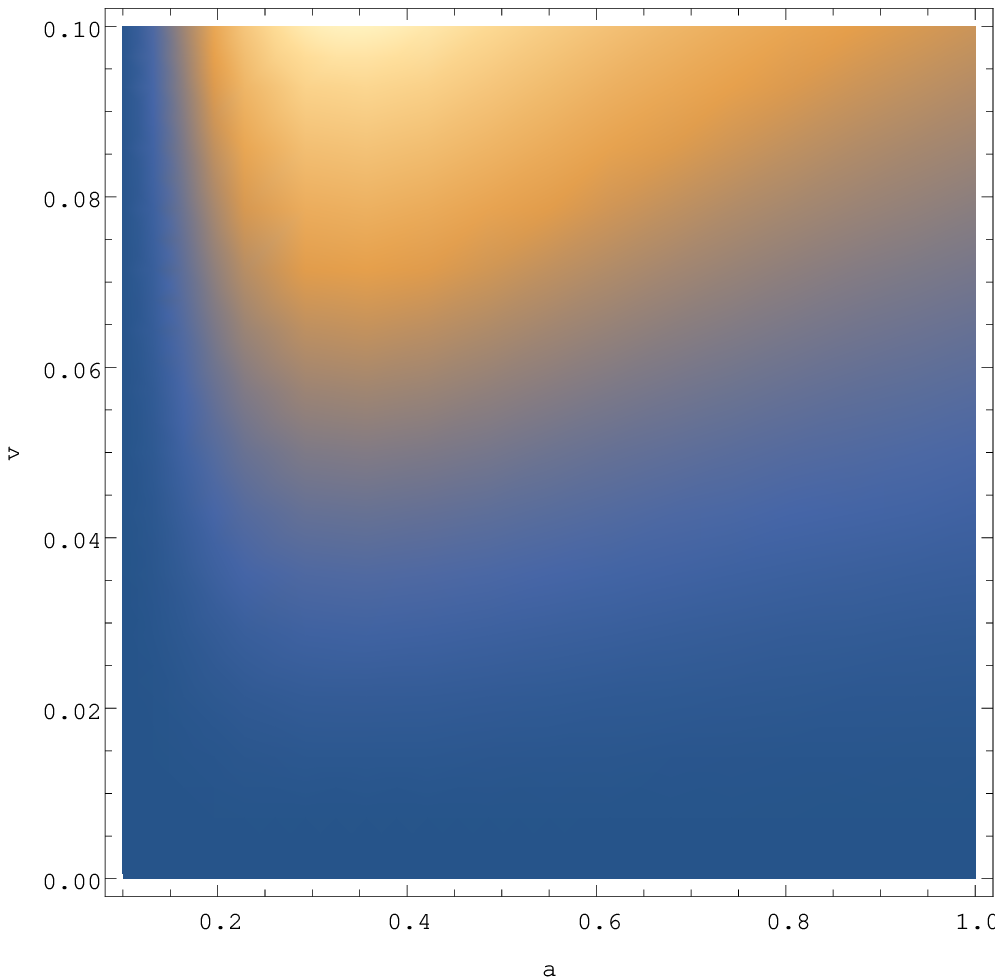}
\caption{The quantum Fisher information for estimating the Unruh temperature $T$ as functions of the acceleration $a$ and the effective coupling parameter $\nu$. The total atoms $n$ and the excited atoms $k$ in the initial state of multiparty system are fixed as $n=3$ and $k=1$, respectively. The energy gap $\Omega$ is fixed  as $\Omega=0.2$. }
\label{QFIDav}
\end{center}
\end{figure}

\begin{figure}
\begin{center}
\includegraphics[height=0.5\textheight]{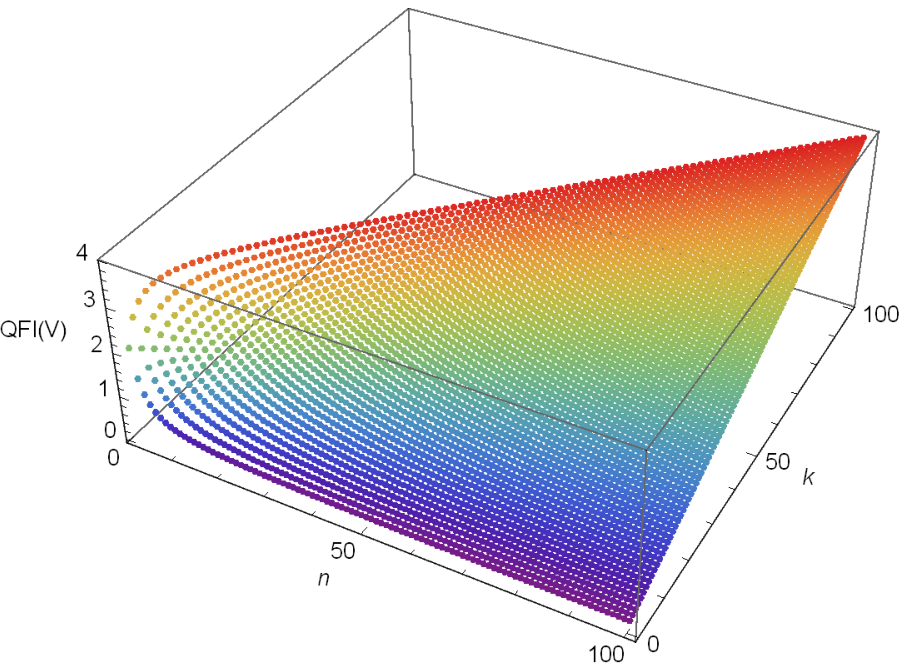}
\caption{ The quantum Fisher information for estimating the  effective coupling parameter $\nu$  versus the total atoms $n$ and the excited atoms $k$ in the initial state of multiparty system. The energy gap $\Omega$, the acceleration $a$ and the  effective coupling parameter $\nu$ are fixed as $\Omega=0.2$, $a=0.3$ and $\nu=0.1$, respectively. }
\label{QFIVnk}
\end{center}
\end{figure}

In Fig. \ref{QFIDav}, the quantum Fisher information for estimating the Unruh temperature $T$  as functions of the acceleration $a$ and the effective coupling parameter $\nu$ is analyzed.
It is show that the quantum Fisher information  is not a monotonic increasing function of  the acceleration $a$. Considering  that  a bigger quantum Fisher information corresponds to a higher precision,  one can select a range of acceleration which provides  better precision for the estimation of the Unruh temperature. Differently, as the effective coupling parameter $\nu$ increases, the quantum Fisher information for estimating the Unruh temperature  increase. That is to say,  the interaction between the atom and the scalar field  would enhance the accuracy for estimating the Unruh temperature in the multiparty clock synchronization protocol.

We also interested in the estimation effective coupling parameter $\nu$  in the accelerated  Unruh-DeWitt detectors. Employing the final state  (\ref{Multiple2}) the definition of quantum Fisher information  (\ref{mixed}), we can also calculate quantum Fisher information for  the  effective coupling parameter $\nu$. After some calculations, the quantum Fisher information for $\nu$ is found to be
\begin{equation}\nonumber
  \mathcal{F}_{Q}(v)=\frac{4n(1-q)[nq+(1-q)k]}{[(1-q)kv^{2}+n(1-q+qv^{2})]^{2}}.\label{quantum Fisher informationT}
\end{equation}
  In Fig. \ref{QFIVnk}, we plot the quantum Fisher information for estimating the effective coupling parameter  versus the total atoms $n$ and the excited atoms $k$ in the initial state of multiparty system. We find that, with more excited atoms in the same initial  probe state, the precision for estimating the  effective coupling parameter is higher. Different from the estimation of Unruh temperature, the estimation of the coupling parameter $\nu$ would gain more accuracy with the $Z$-type state (if $k\neq1$) instead of the $W$-type state. But similar with the estimation of Unruh temperature, if the number of total atoms $n$ is smaller, the quantum Fisher information for estimating the coupling parameter would decrease, which means that a smaller multiparty system doesn't support the estimation of the parameter $\nu$ in the many-body Unruh-DeWitt detector model.

In fact,  the quantum Fisher information is a measure of macroscopic coherence because it can be  demonstrated as the coherence of many copies of a state \cite{Yadin16}.  The macroscopic coherence can be quantified by a superposition of states differing from one another in $A$-value by a fixed amount $\delta$.  For the observable $A$, the quantum Fisher information
 can be expressed as \cite{Yadin16}
\begin{equation} \label{eqn:fisher_expansion}
\mathcal{F}_{Q}(\rho, A) = 2 \sum_{a,b} \frac{(\lambda_a - \lambda_b)^2}{\lambda_a + \lambda_b}|\langle \psi_a | A |\psi_b\rangle |^2,
\end{equation}
where $\rho = \sum_a \lambda_a |\psi_a \rangle \langle \psi_a |$ is a spectral decomposition and the sum is over all $a,\,b$. For pure states, $\mathcal{F}_Q(|\psi\rangle \langle\psi |,A) = 4V(|\psi\rangle,A)$, where $V(|\psi\rangle,A)=|\langle \psi_a | A^2 |\psi_b\rangle |-|\langle \psi_a | A |\psi_b\rangle |^2$. For mixed states,  assuming  $|\psi\rangle$ is a reference state with $V(|\psi\rangle,A)=A_0$, the macroscopic coherence  for $n$ copies can be  defined via the observable $\sum_{i=1}^n A_i$. In the limit of large $n$, $|\psi\rangle^{\otimes n}$ and $|\psi\rangle^{\otimes m}$ have the same macroscopic coherence  for all $\delta$, then we have $m/n = V(|\psi\rangle,A)/A_0$ \cite{Yadin16}. The minimal average ratio $m/n$ over all pure state decompositions $\rho = \sum_\mu p_\mu |\psi_\mu\rangle \langle\psi_\mu |$ is $\mathcal{F}_{Q}(\rho,A)/(4A_0)$ \cite{Yadin16}.  This means that  the macroscopic coherence depends only on this distribution. If  one take the reference state as a product of single-qubit states $\bigotimes_{i=1}^N |\phi_i\rangle$ with $V(|\phi_i\rangle,A_i)=1$, the average ratio of copies is exactly $\mathcal{F}_{Q}(\rho,A)/(4N)$, where $\mathcal{F}_{Q}(\rho,A)$ is the quantum Fisher information  for the  `macroscopic observable' $A$ and N is the number of qubits. That is to say, the quantum Fisher information can be employed to measure the maximum
macroscopic coherence over all observables in $A$ \cite{Yadin16}. Therefore, the results of quantum Fisher information can in principle be used to understandin  the behavior of macroscopic  coherence embedded in the many detector system, which demands later study.

\section{conclusions}
\label{sec:conclusion}
In conclusion, we study the quantum Fisher information, a key concept in quantum metrology, in the estimation of Unruh temperature $T$ and the effective coupling parameter $\nu$ via an entangled many-body system. We find that the precision for the estimation of Unruh temperature in the multiparty entangled probe scheme performs better than the precision in  two-party entangled probe systems. In addition, the precision of estimating Unruh effect would increase when the excited atoms become less in the probe state. It is shown that the proper acceleration for Bob's detector and the interaction between the accelerated detector and the external field have significant influences on the precision for the Unruh effect's estimation. To be specific, there are a range of proper accelerations  that provide us witb a better precision in the estimation of the Unruh temperature. However, one should choose the largest effective coupling  strength  to achieve this goal. Alternatively, we can get a higher precision, i.e., a larger quantum Fisher information for a shorter interaction time and bigger the energy gaps. It is also found that, different from the estimation of Unruh temperature, it requires the $Z$-type initial state with more excited atoms for the estimation of the effective coupling parameter.

\begin{acknowledgments}
This work is supported by the Hunan Provincial Natural Science
Foundation of China  under Grant No. 2018JJ1016; and the National Natural Science Foundation
of China under Grant  No. 11675052 and No. 11875025; and Science and Technology Planning Project of Hunan Province under Grant No. 2018RS3061.		
\end{acknowledgments}

\end{document}